\documentclass[nofootinbib,prb,twocolumn,superscriptaddress,showkeys,preprintnumbers,amsmath,amssymb]{revtex4}

\newif\ifjpg
\jpgtrue

\usepackage{graphicx}
\usepackage{natbib}
\usepackage{color}

\begin{document}

\title{Anisotropic straining of graphene using micropatterned SiN membranes}

\author{Francesca F. Settembrini}
\affiliation{NEST, Istituto Nanoscienze-CNR and Scuola Normale Superiore, I-56126 Pisa,~Italy}

\author{Francesco Colangelo}
\email{francesco.colangelo1@sns.it}
\affiliation{NEST, Istituto Nanoscienze-CNR and Scuola Normale Superiore, I-56126 Pisa,~Italy}
\affiliation{Fondazione Bruno Kessler (FBK), via Sommarive 18, 38123 Povo, Trento, Italy}
 
\author{Alessandro Pitanti} 
\email{alessandro.pitanti@sns.it}
\affiliation{NEST, Istituto Nanoscienze-CNR and Scuola Normale Superiore, I-56126 Pisa,~Italy}

\author{Vaidotas Miseikis}
\affiliation{Center for Nanotechnology Innovation @NEST, Istituto Italiano di Tecnologia, Piazza San Silvestro 12, 56127 Pisa,~Italy}
\affiliation{Istituto Italiano di Tecnologia, Graphene Labs, Via Morego 30, I-16163 Genova,~Italy}

\author{Camilla Coletti}
\affiliation{Center for Nanotechnology Innovation @NEST, Istituto Italiano di Tecnologia, Piazza San Silvestro 12, 56127 Pisa,~Italy}
\affiliation{Istituto Italiano di Tecnologia, Graphene Labs, Via Morego 30, I-16163 Genova,~Italy}
 
\author{Guido Menichetti}
\affiliation{Department of Physics ``E. Fermi'', Universit\`a di Pisa, Largo Pontecorvo 3, I-56127 Pisa, Italy}
\affiliation{NEST, Istituto Nanoscienze-CNR and Scuola Normale Superiore, I-56126 Pisa,~Italy}

\author{Renato Colle} 
\affiliation{DICAM, University of Bologna, via Terracini 28, I-40136 Bologna, Italy}
\affiliation{Department of Physics ``E. Fermi'', Universit\`a di Pisa, Largo Pontecorvo 3, I-56127 Pisa, Italy}

\author{Giuseppe Grosso}
\affiliation{Department of Physics ``E. Fermi'', Universit\`a di Pisa, Largo Pontecorvo 3, I-56127 Pisa, Italy}
\affiliation{NEST, Istituto Nanoscienze-CNR and Scuola Normale Superiore, I-56126 Pisa,~Italy}
 
\author{Alessandro Tredicucci}
\affiliation{Department of Physics ``E. Fermi'', Universit\`a di Pisa, Largo Pontecorvo 3, I-56127 Pisa, Italy}
\affiliation{NEST, Istituto Nanoscienze-CNR and Scuola Normale Superiore, I-56126 Pisa,~Italy}
\affiliation{Fondazione Bruno Kessler (FBK), via Sommarive 18, 38123 Povo, Trento, Italy}
 
\author{Stefano Roddaro}
\affiliation{NEST, Istituto Nanoscienze-CNR and Scuola Normale Superiore, I-56126 Pisa,~Italy}

\begin{abstract}

We use micro-Raman spectroscopy to study strain profiles in graphene monolayers suspended over SiN membranes micropatterned with holes of non-circular geometry. We show that a uniform differential pressure load $\Delta P$ over elliptical regions of free-standing graphene yields measurable deviations from hydrostatic strain conventionally observed in radially-symmetric microbubbles. The top hydrostatic strain $\bar{\varepsilon}$ we observe is estimated to be $\approx0.7\%$ for $\Delta P = 1\,{\rm bar}$ in graphene clamped to elliptical SiN holes with axis $40$ and $20\,{\rm \mu m}$. In the same configuration, we report a $G_\pm$ splitting of $10\,{\rm cm^{-1}}$ which is in good agreement with the calculated anisotropy $\Delta\varepsilon \approx 0.6\%$ for our device geometry. Our results are consistent with the most recent reports on the Gr\"uneisen parameters. Perspectives for the achievement of arbitrary strain configurations by designing suitable SiN holes and boundary clamping conditions are discussed.

\end{abstract}

\keywords{graphene, strain, micro-Raman}

\maketitle

Graphene displays a range of remarkable properties that have catalyzed -- since its discovery in 2004~\cite{novoselov2005two} -- an impressive interest in the scientific community~\cite{geim2007rise}. Its unique electronic behavior stems from the hexagonal honeycomb structure of the carbon lattice, which forces low-energy conducting electrons to assume a linear dispersion that mimicks massless relativistic fermions~\cite{neto2009electronic}. In addition, graphene displays an unusual mechanical strength, and strains up to beyond $10\%$ can be applied without damaging appreciably its structure~\cite{lee2008measurement}. This feature, combined with its intrinsic two-dimensional nature opens unique perspectives for the investigation of strain engineering~\cite{pereira2009strain,low2010strain,guinea2012strain} and for the development of novel device concepts~\cite{qi2013resonant}. In fact, it has been predicted~\cite{geim2009graphene,guinea2012strain}, and in part experimentally demonstrated~\cite{levy2010strain,polini2013artificial}, that mechanical deformations in graphene can be used to tailor its electron properties. As a particularly inspiring possibility, it is known that a suitable deformation of the honeycomb lattice can be equivalent to the application of a pseudomagnetic field~\cite{low2010strain,guinea2010energy}.

\begin{figure}[ht!]
\ifjpg
	\includegraphics[width=0.46\textwidth]{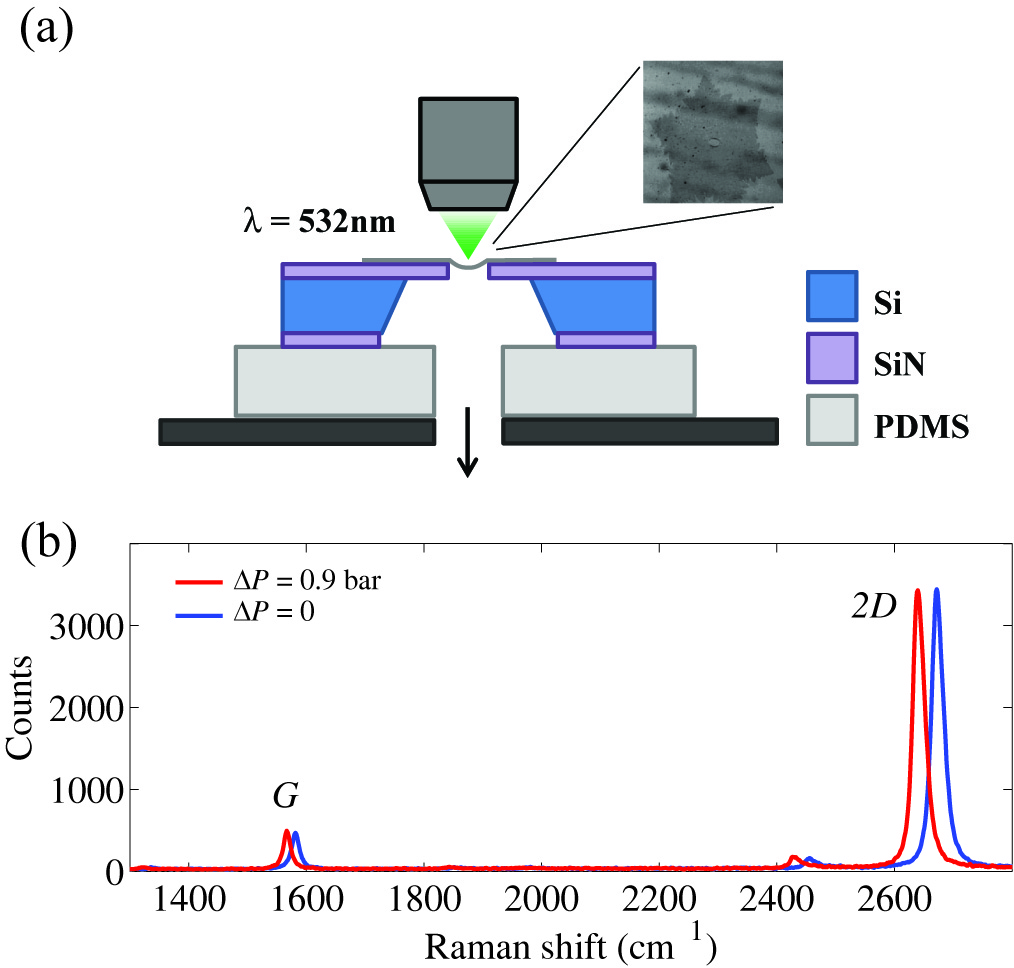}
\else
	\includegraphics[width=0.46\textwidth]{Figure1.pdf}
\fi
\caption{{\bf Straining graphene with a differential pressure load.} (a) Sketch of the device architecture. Monolayer CVD graphene is transferred on a patterned SiN membrane. The bottom of the chip is coupled to a vacuum chamber using a polydimethylsiloxane layer. Deformed graphene is investigated by micro-Raman spectroscopy as a function of the applied differential pressure $\Delta P$. Right picture: optical image of one of the CVD monocrystals deposited on the patterned Si$_3$N$_4$.(b) Raman spectrum measured at $\Delta P=0$ (blue curve) and $\Delta P=0.9\,{\rm bar}$ (red curve) for an elliptical SiN membrane with axes $20\,{\rm \mu m}$ and $10\,{\rm \mu m}$. }
\end{figure} 

Achieving a controlled strain profile in graphene poses non-trivial technical challenges and various alternative approaches have been explored during recent years. Hydrostatic strain configurations were obtained using circular holes and a uniform differential pressure load~\cite{shin2016raman,zabel2012raman}. In this device architecture, local strain was also explored taking advantage of scanning probe techniques~\cite{zhu2014pseudomagnetic}. Concerning uniaxial strain, various studies have demonstrated the possibility to anisotropically deform graphene deposited on polidimethylsiloxane (PDMS)~\cite{huang2009phonon} or on similar stretchable substrates~\cite{mohiuddin2009uniaxial}. An alternative promising approach consists in anchoring graphene layers to micro-electromechanical actuators~\cite{perez2014controlled}. More elaborated strain profiles, in particular those giving rise to a pseudomagnetic field, have been hard to demonstrate so far. Interesting experimental evidences have been put forward in the context of random nanobubbles~\cite{levy2010strain} and fascinating results have been obtained in deformed artificial honeycomb structures mimicking the behavior of graphene~\cite{polini2013artificial}. In practice, the achievement of custom strain profiles has generally proved to be rather elusive.

In the present work, we demonstrate that non-trivial strain profiles can be obtained in free-standing graphene membranes that are clamped on an edge that is not radially symmetric and are subject to a vertical uniform load using a pressure difference between the two opposite faces of the graphene flake. In particular, we show that loaded elliptical membranes display Raman features that demonstrate the presence of an anisotropic component in the induced strain profile, in good agreement with what expected with the studied geometry. This proof of principle demonstration delineates a novel strategy for achieving and controlling complex strain profiles in graphene, based on the design of custom clamping geometries for free-standing graphene flakes.

Figure~1 shows the device architecture and set-up adopted for this work. Free-standing graphene areas of various shapes and dimensions were obtained using micropatterned SiN membranes as the mechanical support for the graphene layer. Starting from a Si wafer doubly coated in ``pre-stressed'' $300\,{\rm nm}$ of Si$_3$N$_4$, a combination of dry and wet etching protocols (see Supplementary Information for further details) were adopted to obtain suspended $500\times500\,{\rm \mu m^2}$ Si$_3$N$_4$ membranes with through holes of various geometries. In the present study, we investigated a set of elliptical holes with various major ($a$) and minor ($b$) axes: $a\times b =5\,{\rm \mu m}\times10\,{\rm \mu m}$, $10\,{\rm \mu m}\times20\,{\rm \mu m}$ and $20\,{\rm \mu m}\times40\,{\rm \mu m}$. Circular holes were also investigated, as reported in the literature~\cite{si2016strain,zabel2012raman}. Large-scale monocrystalline graphene flakes used in the present work were obtained by CVD growth on Cu~\cite{miseikis2015rapid} and transferred on the Si$_3$N$_4$/Si chips using a standard ``bubbling transfer'' technique~\cite{gao2012repeated}. As sketched in Fig.~1a, a differential pressure $\Delta P$ is applied orthogonally to the free-standing graphene region thanks to a $2\,{\rm mm}$-thick polydimethylsiloxane (PDMS) coupling layer placed on top of a modified microscope slide. As a result, the top side of the free-standing graphene region is subject to ambient pressure (conventionally $P_0=1\,{\rm bar}$) while the bottom space can be partially or completely evacuated with a scroll pump. Static vacuum tests were performed to verify the stability of $\Delta P$, which was found to decay over a time scale of various hours: this ensures pressure values measured by the gauge are meaningful. Maps were always performed under active pumping conditions, at $\Delta P=1\,{\rm bar}$.

\begin{figure}[h!]
\ifjpg
	\includegraphics[width=0.48\textwidth]{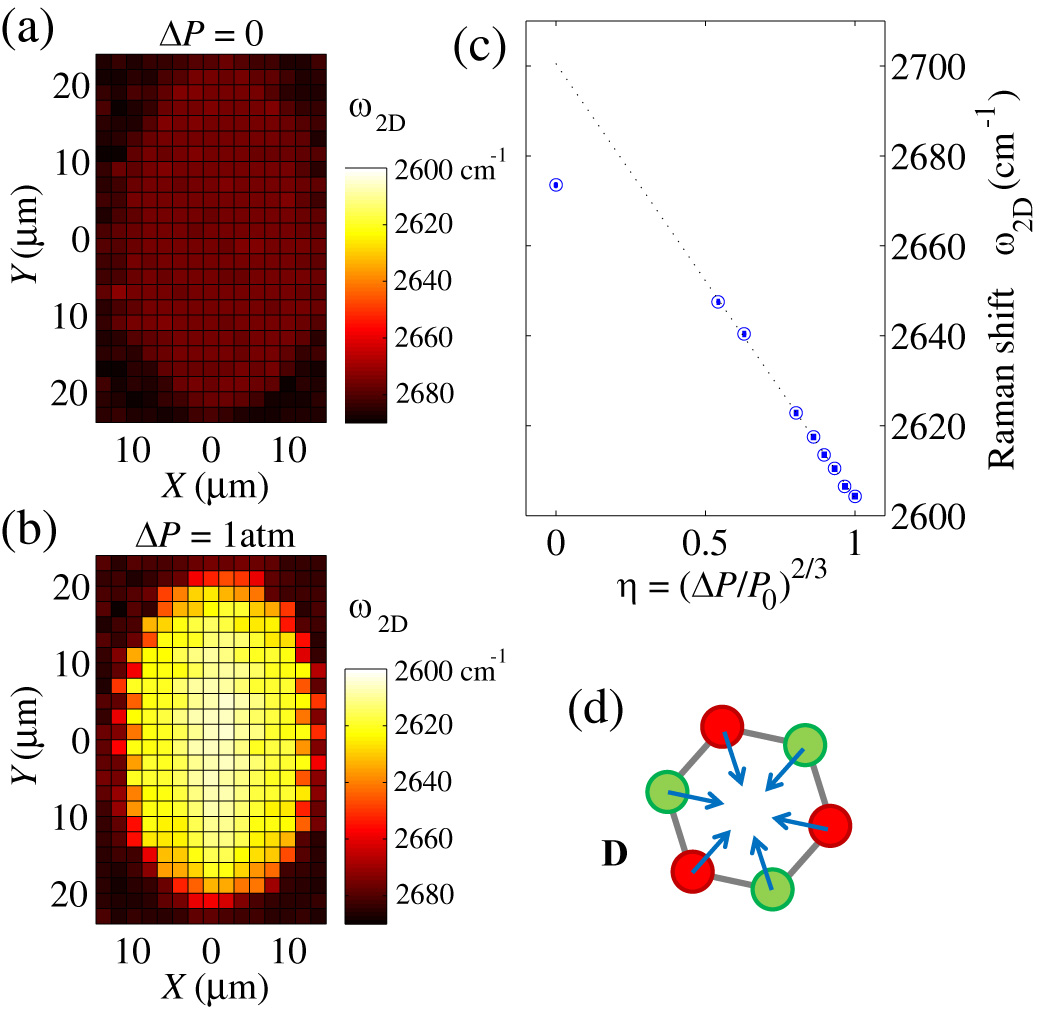}
\else
	\includegraphics[width=0.48\textwidth]{Figure2.pdf}
\fi
\caption{{\bf Impact of strain on the 2D peak.} (a) Map of the position of the 2D peak for $\Delta P=0$: the peak position is mostly uniform with a slight red shift over the suspended region (see discussion in the main text). (b) The application of a differential pressure $\Delta P=1\,{\rm bar}$ leads to a dome-shaped shift, which is maximal at the center of the membrane. As further argued based on the data presented in Fig.~3, this effect can be explained as the consequence of a hydrostatic strain $\bar{\varepsilon}$ in the elliptical hole. (c) Evolution of the Raman shift as a function of the parameter $\eta = (\Delta P/P_0)^{2/3}$, which is proportional to $\bar{\varepsilon}$. A linear regression of the observed Raman shift (excluding the value at $\eta = 0$) and a comparison with numerical estimates of the strain yields a Gr\"uneisen parameter consistent with the most recent results reported in the literature. (d) Sketch of the $D$ mode in graphene, whose second order causes the 2D Raman resonance.}
\end{figure} 

Local graphene deformation is investigated by micro-Raman spectroscopy, using an {\em inVia} confocal system by Renishaw equipped with a polarized $\lambda = 532\,{\rm nm}$ laser source. The Raman signal was collected through a 50X objective with N.A.=0.75 and analyzed by a 1800 grooves/mm grating. Raman maps were collected using a laser power of $1\,{\rm mW}$ to minimize the impact of local heating. In Fig.~1b we report the measured Raman spectra collected at the center of a $10\,{\rm \mu m}\times 20\,{\rm \mu m}$ elliptical graphene region, for $\Delta P=0$ (blue curve) and $\Delta P=0.9\,{\rm bar}$ (red curve): as expected, the $G$ and $2D$ Raman peaks are significantly red shifted by the strain in the suspended graphene region. Importantly, the modification of the Raman spectra was always found to be completely reversible upon removal of the pressure load: this proves that no measurable adjustment nor sliding of graphene occurs during our experiments.

\begin{figure*}[t!]
\ifjpg
	\includegraphics[width=0.95\textwidth]{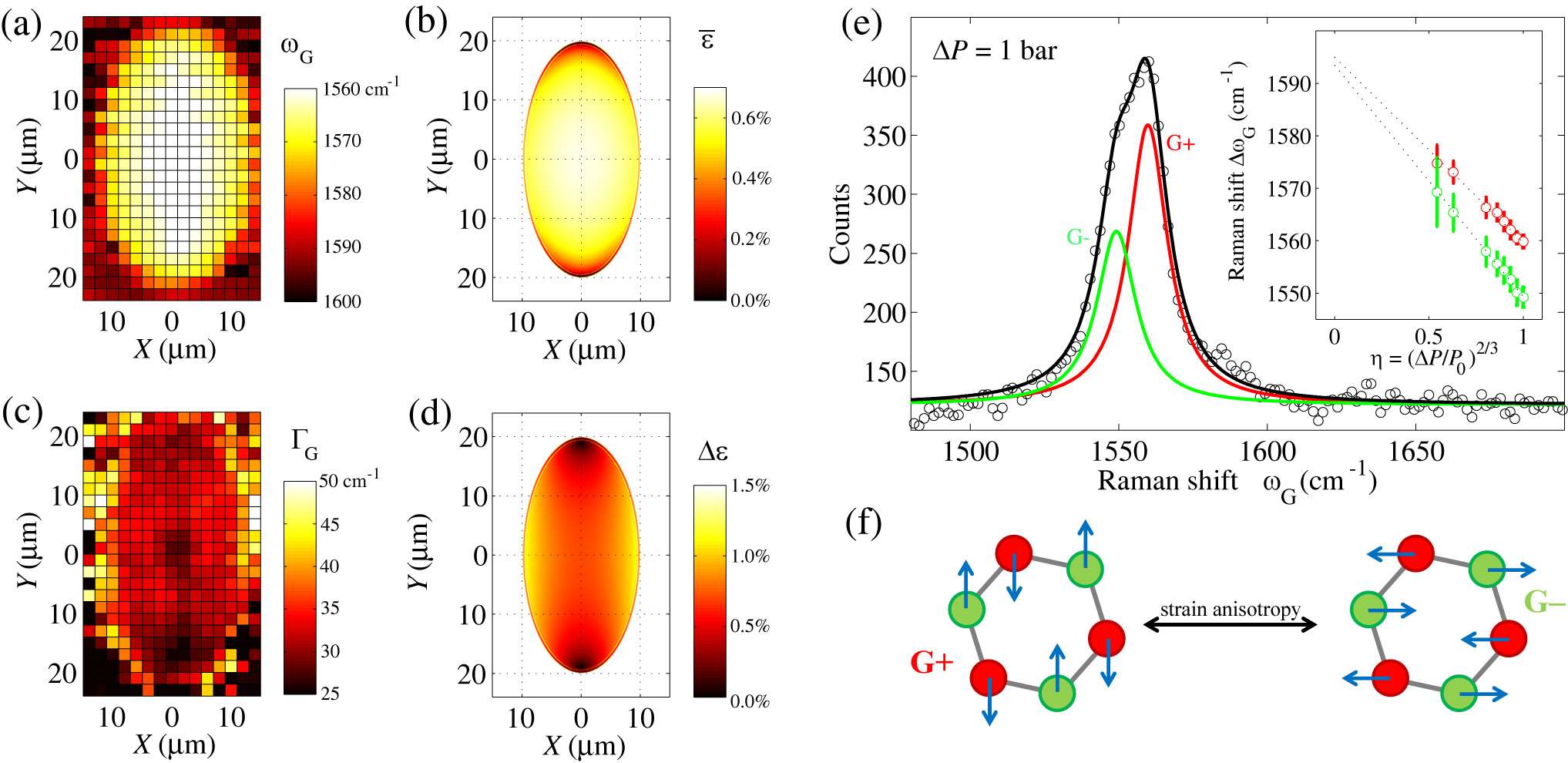}
\else
	\includegraphics[width=0.95\textwidth]{Figure3.pdf}
\fi
\caption{{\bf Strain-induced shift and splitting of the G peak.} (a) Map of the Raman shift $\omega_G$ obtained by fitting the $G$ peak with a single Lorentzian curve. (b) Simulated average strain map $\bar{\varepsilon} = (\varepsilon_{xx}+\varepsilon_{yy})/2$ at $\Delta P=1\,{\rm bar}$. (c) Map of the peak broadening $\Gamma_G$ obtained by fitting the $G$ peak with a single Lorentian curve. (d) Simulated strain anisotropy map $\Delta\varepsilon = \sqrt{(\varepsilon_{xx}-\varepsilon_{yy})^2+4\varepsilon_{xy}^2}$ at $\Delta P=1\,{\rm bar}$. (e) Multi peak fit of the $G$ peak at $\Delta P=1\,{\rm bar}$ measured at the center of the elliptical hole. The resulting positions of the $G_+$ (red curve and markers) and $G_-$ (green curve and markers) peak versus $\Delta P$ are reported in the inset along with a weighted linear fit. (f) Sketch of the $G_+$ and $G_-$ modes in the presence of an arbitrary anisotropic strain.}
\end{figure*} 

The most evident impact of deformation is visible in Fig.~2, where we compare the map of the $2D$ peak Raman shift $\omega_{2D}$ for $\Delta P=0$ (Fig.~2a) and $\Delta P=1\,{\rm bar}$ (Fig.~2b). For symmetry reason its center frequency is sensitive to the hydrostatic component of the strain tensor $\varepsilon_{ij}$, which we name $\bar{\varepsilon} = (\varepsilon_{xx}+\varepsilon_{yy})/2$. Experimentally, the $2D$ peak displays a maximal red shift at the center of the suspended region, similarly to what reported for inflated circular graphene membranes~\cite{zabel2012raman}. The quantitative evolution of the shift versus $\Delta P$ is highlighted in Fig.~2c where we report $\omega_{2D}$ measured at the maximal shift region in the center of the ellipse. The phononic mode giving rise to the higher order $2D$ peak is sketched in Fig.~2d. Raman shifts at various pressure loads are compared with $\eta = (\Delta P/P_0)^{2/3}$, where $P_0=1\,{\rm bar}$: all the components of the strain tensor $\varepsilon_{ij}$ at the center of the membrane are in fact expected to scale linearly with the power $2/3$ of the pressure load (see Supplementary Information). The overall dependence of $\omega_{2D}$ as a function of $\Delta P$ and of the position on the ellipse is found to be largely consistent with reports on the simpler case of radially symmetric graphene clamping and with the most recent estimates of the Gr\"uneisen parameters~\cite{shin2016raman,mohiuddin2009uniaxial,cheng2011gruneisen}. We also note that, consistently with reported data~\cite{shin2016raman}, the value of $\omega_{2D}$ for $\Delta P\approx 0$ is found to display a further surprising red shift. This effect was previously attributed to uncertainties in the exact determination of $\Delta P$~\cite{shin2016raman}. We believe that an additional reason for the shift is possibly related to graphene adhesion on the vertical sidewalls of the SiN hole, which is known to occur in this kind of graphene drums~\cite{shin2016raman} and could be relevant in the low-$\Delta P$ regime. The verification of this hypothesis will likely require a combined Raman and atomic force microscopy study at low pressure loads, that goes beyond the scope of the present work. Further details regarding the numerical calculation of the $\varepsilon_{ij}$ tensor as a function of $\Delta P$ and scaling rules are reported in the Supplementary Information.

While hydrostatic deformations explain well the coarse evolution of the Raman spectra, the strain profiles in our elliptically clamped graphene membranes are expected to display a marked deviation from a uniform strain configuration and a larger strain can be expected along the shorter axis of the ellipse. More in general, the anisotropic component of the strain field can be expressed through the invariant

\begin{align}
\Delta\varepsilon = \sqrt{(\varepsilon_{xx}-\varepsilon_{yy})^2+4\varepsilon_{xy}^2}
\end{align}

\noindent corresponding to the difference between two eigenvalues of the strain tensor $\varepsilon = \bar{\varepsilon}\pm\Delta\varepsilon/2$. It is well known~\cite{cheng2011gruneisen} that strain anisotropy, when sufficiently large, can be detected in Raman spectroscopy as a splitting of the degenerate phononic modes $G_\pm$. In Fig.~3 we report a detailed study of the Raman spectrum of the $G$ peak region as a function of $\Delta P$. Data reported in Fig.~3 refer to the largest explored Si$_3$N$_4$ elliptical $20\times 40\,{\rm \mu m^2}$ hole: larger suspended areas in fact correspond -- for a given value of $\Delta P$ -- to a larger anisotropic strain $\Delta\varepsilon$.

A first rough analysis was performed by fitting the Raman data with a single lorentzian peak. As visible in Fig.~3a, the resulting map of the Raman shift $\omega_G$ at $\Delta P=1\,{\rm bar}$ is found to be consistent with the $\omega_{2D}$ map reported in Fig.~2a. For comparison, we report in Fig.~3b the map of $\bar{\varepsilon}$ calculated for the same pressure load. A top hydrostatic strain of $0.68\%$ is expected, in agreement with the observed red shift of the $G$ peak and known values of the corresponding Gr\"uneisen parameter (see Supplementary Information for further details). As argued in the following, on the other hand, hydrostatic strain is not sufficient to satisfactorily describe the evolution of the $G$ peak as a function of the pressure load. Indeed, a non-trivial broadening is visible in Fig.~3c, where we report the FWHM $\Gamma_G$ resulting from the same fitting procedure. The observed broadening displays a peculiar ``saddle point'' spatial evolution, with a $\Gamma_G$ broadening which is significantly smaller than the average at the top and bottom apexes of the elliptical hole. Remarkably, a very similar pattern is visible in Fig.~3d, displaying the calculated $\Delta\varepsilon$ for the same clamping geometry, at $\Delta P=1\,{\rm bar}$. This suggests that, beyond mechanisms highlighted in recent works~\cite{shin2016raman}, broadening in our experiment is also connected to strain anisotropy.

A more detailed investigation of the broadening mechanism was performed through the analysis of the $G$ peak measured at the center of the elliptical hole, which represents a good trade-off between the expected value of $\Delta\varepsilon$ and the minimization of the impact of the borders of the Si$_3$N$_4$ hole. In this position, numerical estimates indicate that a top anisotropic strain component $\Delta\varepsilon = 0.64\%$ can be expected. The Raman spectrum in the $G$ peak region for $\Delta P=1\,{\rm bar}$ is reported in Fig.~3e: the peak displays a lineshape which is clearly consistent with the superposition of two nearby lorentzian peaks, which we interpret as corresponding to the $G_+$ and $G_-$ modes in uniaxially strained graphene (see Fig.~3f). A similar analysis (see Supplementary Information for further information concerning the fitting procedure) was performed for various values of $\Delta P$ and the resulting peak positions are reported in the inset to Fig.~3e. Two divergent peaks are obtained with a top splitting of about $10\,{\rm cm^{-1}}$, which is in very good agreement~\cite{cheng2011gruneisen} with what expected for an anisotropy $\Delta\varepsilon=0.64\%$. A weighted linear regression of the two peak positions $\omega_{G_{\pm}}$ yields two remarkably linear trends in $\eta$ that converge almost exactly for $\Delta P=0$, as expected.

Our work demonstrates that non-trivial strain profiles can be obtained in Si$_3$N$_4$ holes with an elliptical shape, and that a sizable anisotropic component in the graphene strain can be obtained. Similarly, we expect that more in general the clamping geometry can be used to design even more advanced non-uniform and non-isotropic strain profiles, taking advantage of a relatively robust implementation with no free graphene edges. In view of the possibility to impact the electronic states via the engineering of custom strain profiles, the observed $G_\pm$ splitting phenomenology has also been compared with a first-principle calculation on an atomistic model system mimicking the experimental setup. To this end, we have simulated an unstrained suspended graphene layer with an elliptic-shape depression (see inset of Fig.~4). To reproduce the experimental configuration, we have fixed the position of the carbon atoms external to the ellipse to a ``zero'' height, as it happens for graphene on Si$_3$N$_4$ substrate, and the effect of the vertical load has been reproduced by fixing the two carbon atoms at the center of the ellipse (blue dots in the inset) at a lower vertical position and leaving all the other carbon atoms in the ellipse free to relax in order to reach the minimum energy structure. The simulation has been performed on a $7.4\,{\rm\AA}\times12.8\,{\rm\AA}$ ellipse containing a total of $22$ carbon atoms. The Raman spectra of the system have been calculated by means of density functional perturbation theory~\cite{baroni2001phonons} as implemented in QUANTUM-ESPRESSO code~\cite{giannozzi2009quantum}, with  local density approximation and  norm-conserving pseudopotential for the carbon atoms~\cite{troullier1991efficient}. We used a plane wave expansion up to $80\,{\rm Ry}$ cutoff and   $4\times4\times1$ Monkhorst-Pack mesh~\cite{monkhorst1976special}  for  the sampling of the Brillouin zone. The ellipse depression was created in the central part of a $5\times5$ graphene supercell with lattice parameter $a_0 =12.30\,{\rm\AA}$.

In the absence of strain, the frequency of the degenerate $G$ phononic mode is $\omega_{G}=1555\,{\rm cm^{-1}}$. The numerical result is smaller than the one experimentally observed for graphene not subject to pressure loads, as likely due to doping and to the interaction with the Si$_3$N$_4$ substrate. For $\Delta P \neq 0$, the value of the effective strain along the ellipse axes has been estimated from the depth of the two central carbon atoms, $\delta$, and the length of the principal axes of the ellipse, $a$ and $b$: e.g. for the major axis we have $\varepsilon_a \sim (2L-a)/a$ where $L \sim \sqrt{(a/2)^2+\delta^2}$ is the profile length for the depression along the $a$ direction; the same holds for the minor axis. The results of the calculation are shown in Fig.~4: the asymmetry of the strain due to the elliptical geometry of the depression is responsible of an averaged splitting of the $G$ peak mode, that is found to be of the same order of magnitude as the one experimentally induced  by uniaxial strain on the graphene layer~\cite{mohiuddin2009uniaxial}.
 
\begin{figure}[ht!]
\ifjpg
	\includegraphics[width=0.46\textwidth]{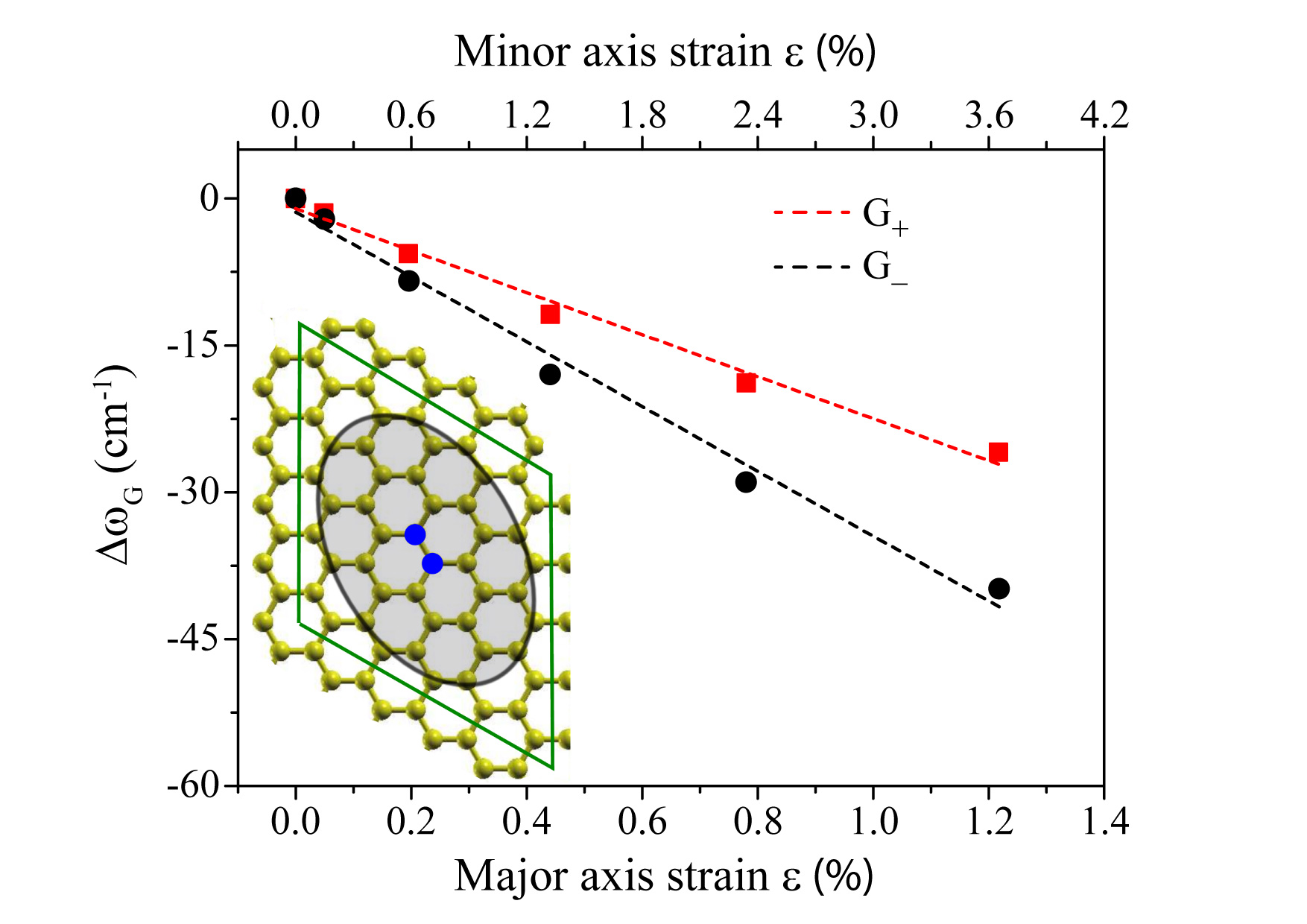}
\else
	\includegraphics[width=0.46\textwidth]{Figure4.pdf}
\fi
\caption{Position of the $G_{\pm}$ peak as function of the major axis strain. The slope of the Raman shifts are $\partial\omega_{G_-}/\partial\varepsilon_a\sim-33\,{\rm cm^{-1}}/\%$ for $G_-$ and  $\partial\omega_{G_+}/\partial\varepsilon_a\sim-21\,{\rm cm^{-1}}/\%$ for $G_+$. The inset shows the simulation cell (green) containing the  ellipse.  The lowest fixed carbon atoms are indicated in blue. }
\end{figure}

In conclusion, we have provided evidence of an incipient splitting of the $G$ mode in free-standing graphene regions clamped to an elliptical hole in Si$_3$N$_4$ and subject to a uniform differential pressure load. Our results indicate a promising route to induce custom strain profiles, which can be controlled by the applied pressure load and {\em designed} according to the chosen geometry of the supporting Si$_3$N$_4$ frame. We also highlight that our experiment has been performed using large scale CVD monocrystalline graphene flakes, thus providing a route for scalable strain-engineered graphene devices. Finally, we would like to stress that present results have been obtained using a vacuum chamber to induce a maximal load $\Delta P=1\,{\rm bar}$. A recent report~\cite{shin2016raman} demonstrates that using pressurized gas as a load it is possible to reach $\Delta P=14\,{\rm bar}$, potentially leading to significantly increased achievable strain magnitude (about a factor six larger strain can be expected) and/or to less stringent limits on the minimal area of the Si$_3$N$_4$ holes.

This work was supported by the EC under the Graphene Flagship program (contract no. CNECT-ICT-604391) and by the ERC advanced grant SoulMan (G.A. 321122). SR acknowledges the support of the CNR through the bilateral CNR-RFBR projects.

\bibliographystyle{myownbib}
\bibliography{mybib}  

\newpage

\renewcommand{\thefigure}{S\arabic{figure}}
\setcounter{figure}{0}

{\bf \large Supplementary Information}

\appendix

\section{Numerical calculation of the strain profiles}

Strain profiles for graphene clamped on an elliptical boundary were calculated using the finite element software COMSOL, using a Young modulus $E=1\,{\rm TPa}$, a Poisson ratio $\nu=0.165$ and assuming the conventional thickness $h=0.335\,{\rm nm}$\footnote{C. Lee, X. Wei, J. W. Kysar, et al. ``Measurement of
the elastic properties and intrinsic strength of monolayer graphene''. {\em Science}, 321, 385–388, (2008).}. Results were studied as a function of the mesh density to rule out spurious effects. Interestingly, the simulation result does not depend in any significant way on the number of mesh layers in the vertical direction and the same results are obtained even using only a {\em single layer} mesh. This indicates that the simulated configurations almost exclusively depend on the elastic stretching of the membrane in the plane direction, while the bending modulus does not appear to have any measurable impact in the simulated configurations. This is relevant to our experiment, since it is well-known that graphene -- being a strongly anysotropic material -- displays a bending modulus which is {\em different} from (in particular, {\em smaller} than) the value $D=Eh^3/12(1-\nu^2)$, which can be calculated assuming three-dimensional isotropic mechanical properties and the conventional thickness $h$. The fact that our numerical result are only sensitive to planar deformations ensure that they are not affected by mechanical anisotropy of graphene.

\begin{figure}[h!]
\ifjpg
	\includegraphics[width=0.48\textwidth]{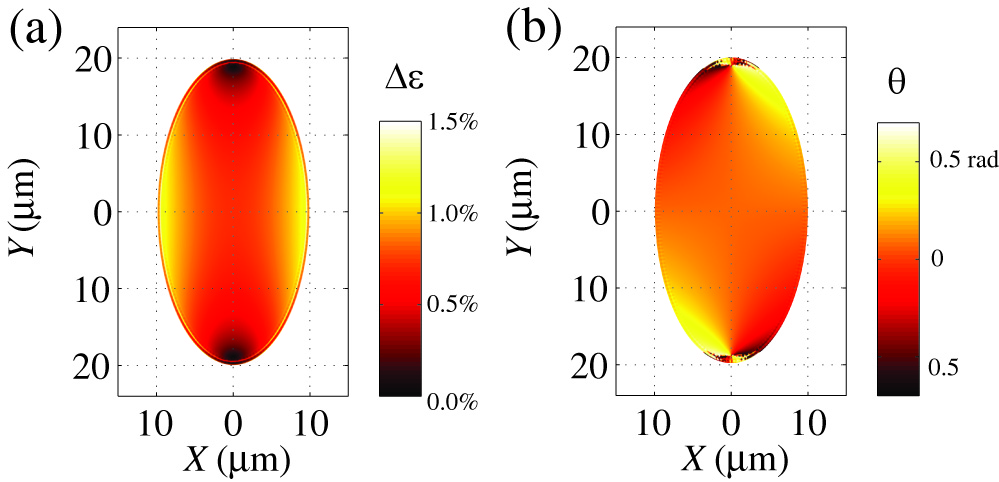}
\else
	\includegraphics[width=0.48\textwidth]{SF1.pdf}
\fi
\caption{{\bf Expected strain anisotropy on elliptical membranes.} (a) Map of the strain anisotropy $\Delta\varepsilon$ for a load $\Delta P=1\,{\rm bar}$. (b) Map of the angle $\theta$ between the $\hat{x}$ axis and the main strain axis of the deformed graphene membrane. In regions where $\Delta\varepsilon$ is large we obtain $\theta\approx0$ and thus the strain anisotropy is in good approximation always directed along the $\hat{x}$ axis.}
\end{figure} 

The resulting $\varepsilon_{ij}$ tensor was thus decomposed into its hydrostatic and anisotropic component $\varepsilon_{ij}=\bar{\varepsilon}\delta_{ij}+\delta\varepsilon_{ij}$, where $\bar{\varepsilon}$ is the average strain. The two-dimensional deviatoric strain $\delta\varepsilon_{ij}$ will have two eigenvalues $\pm\Delta\varepsilon/2$, where $\Delta\varepsilon$ is the strain anisotropy, i.e. the difference between the strain eigenvalues of $\varepsilon_{ij}$. It should be noted that the anisotropy is mainly, but not exacly directed along the minor axis of the elliptical hole. In Fig.~S1, we report the map of $\Delta\varepsilon$ (panel a) and the direction of the main strain axis (panel b). In the regions where $\Delta\varepsilon$ is large, the main strain axis has an almost constant direction coincident with the $\hat{x}$ axis. The main strain axis was calculated according to 

\begin{align}
\tan 2\theta = \frac{2\varepsilon_{xy}}{\varepsilon_{xx}-\varepsilon_{yy}}.
\end{align}

The evolution of the strain magnitude in general, and of the strain anisotropy in particular, as a function of the pressure load $\Delta P$ is analyzed in Fig.~S2. For thin circular membranes, it is well known\footnote{H. Hencky. ``Uber den Spannungszustand in kreisrunden Platten mit verschwindender Biegungssteifigkeit''. {\em Z. Math. Phys.}, 63, 311–317, (1915).}$^,$\footnote{U. Komaragiri, M. Begley, and J. Simmonds. ``The mechanical response of freestanding circular elastic films under
point and pressure loads''. {\em Journal of Applied Mechanics}, 72, 203–212, (2005).} that the maximum displacement $\delta$ at the center of the membrane scales at $\delta\propto \Delta P^{1/3}$. This, combined with simple assumptions about the shape of the deformed membranes\footnote{Y. Shin, M. Lozada-Hidalgo, J. L. Sambricio, et al. ``Raman spectroscopy of highly pressurized graphene membranes''. {\em Applied Physics Letters}, 108, 221907, (2016).}$^,$\footnote{J. S. Bunch, S. S. Verbridge, J. S. Alden, et al. ``Impermeable atomic membranes from graphene sheets''. {\em Nano Letters}, 8, 2458–2462, (2008).} leads to a strain tensor which satisfies $\varepsilon_{ij}\propto \Delta P^{2/3}$. Our numerical calculations indicate that the same dependence holds for elliptical membranes. In panel (a), we report the value of the vertical displacement at the center of an elliptical $40\,{\rm \mu}\times 20\,{\rm \mu m}$ membrane, as a function of the parameter $(\Delta P/P_0)^{1/3}$ and confirm a linear dependence between the two. In panel (b), we report the value of $\varepsilon_{xx}$ and $\varepsilon_{yy}$. We note that at the center of the ellipse $\varepsilon_{xy}=0$ for symmetry reasons and thus $\bar{\varepsilon}=(\varepsilon_{xx}+\varepsilon_{yy})/2$ and $\Delta\varepsilon = |\varepsilon_{xx}-\varepsilon_{yy}|$. All the strain parameters are found to depend, up to a very good approximation, linearly on $\eta=(\Delta P/P_0)^{2/3}$.

\begin{figure}[h!]
\ifjpg
	\includegraphics[width=0.46\textwidth]{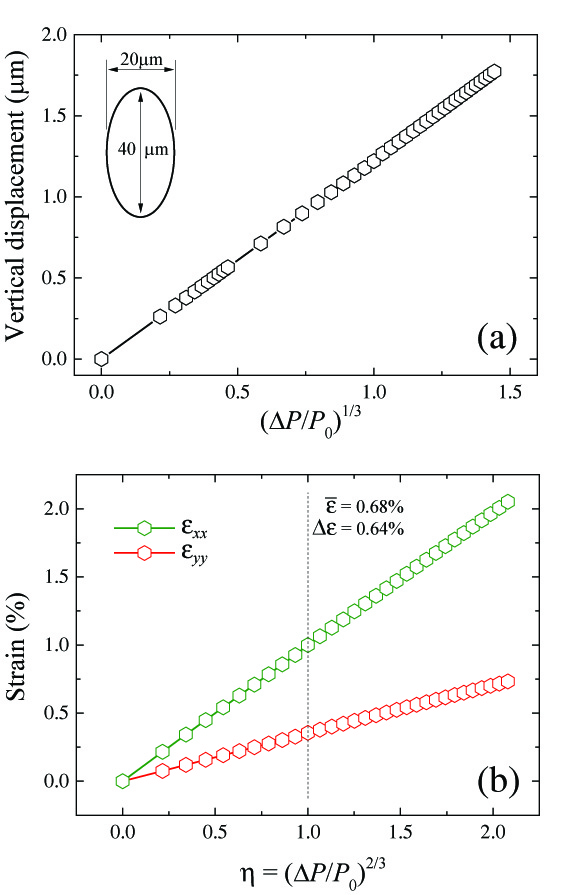}
\else
	\includegraphics[width=0.46\textwidth]{SF2.pdf}
\fi
\caption{{\bf Dependence of strain versus the pressure load.} (a) The calculated vertical displacement at the center of the ellipse (axes $40\,{\rm \mu m}\times20\,{\rm \mu m}$) is found to be proportional to $(\Delta P/P_0)^{1/3}$, consistently with known results on circular membranes. (b) The corresponding strain components $\varepsilon_{xx}$ and $\varepsilon_{yy}$ ($\varepsilon_{xy}=0$ at the center of the membrane) are shown: both the components, and thus also $\bar{\varepsilon}$ and $\Delta\varepsilon$, are proportional to $(\Delta P/P_0)^{2/3}$. The values obtained at $\eta=1$ are $\bar{\varepsilon}=0.68\%$ and $\Delta\varepsilon = 0.64\%$.}
\end{figure} 

\section{Gr\"uneisen parameters}

Based on our numerical results, the experimental Raman shifts can be used to extract an estimate of the Gr\"uneisen parameters for the $G$ and $2D$ resonances in graphene. Experimentally, both $\Delta\omega_G$ and $\Delta\omega_{2D}$ shift linearly in $\eta=(\Delta P/P_0)^{2/3}$, except for $\Delta P \approx 0$, where they are typically found to display a further red shift and to slightly depart from linear dependence. A similar effect was reported in recent literature and it was attributed to uncertainties in the determination of the exact value of $\Delta P$. We argue that a possible further source of the effect could also be related to the pre-stress of the graphene membrane due to adhesion to the vertical walls of the SiN hole\footnote{C. Lee, X. Wei, J. W. Kysar, et al. ``Measurement of
the elastic properties and intrinsic strength of monolayer graphene''. {\em Science}, 321, 385–388, (2008).}. This effect is expected to be relevant for low values of $\Delta P$, while large pressure loads can be expected to lead to a detachment of graphene from the sidewalls. We notice that, together with substrate doping, sidewall adhesion could also cause the red-shift of the Raman peaks observed on suspended graphene with respect to the SiN substrate (see Fig.~2a in the main text). The impact of adhesion is expected to be visible in atomic force microscopy as a function of $\Delta P$, but such a test could not be performed in the present experiment. This deviation was disregarded in our analysis and all linear fits were performed excluding the Raman shift value at $\Delta P=0$.

Based on our numerical results, we obtain for $\eta=1$ an average strain $\bar{\varepsilon} = 0.68\%$ and $\Delta\varepsilon = 0.64\%$. Using this calibration, we extract the following peak shift parameters: $-134\,{\rm cm^{-1}/\%}$ for the 2D peak; $-58.6\pm3.2\,{\rm cm^{-1}/\%}$ for the average between the $G_+$ and $G_-$ peak positions; $13.8\pm0.8\,{\rm cm^{-1}/\%}$ for the $G$ peak splitting. This yields to $\gamma_G=1.83\pm0.10$, $\beta_G=0.87\pm0.05$ and $\gamma_{2D}=2.64\pm0.11$. Errors were calculated based on a $40\,{\rm mbar}$ uncertainty over the value of $\Delta P$ and on the consequent fit errors. Our results are compatible with recent works on strained graphene\footnote{Y. Cheng, Z. Zhu, G. Huang, et al. ``Gr\"uneisen parameter of the G mode of strained monolayer graphene''. {\em Physical Review B}, 83, 115449, (2011).}$^,$\footnote{T. Mohiuddin, A. Lombardo, R. Nair, et al. ``Uniaxial strain in graphene by Raman spectroscopy: G peak splitting, Gr\"uneisen parameters, and sample orientation''. {\em Physical Review B}, 79, 205433, (2009).}$^,$\footnote{Y. Shin, M. Lozada-Hidalgo, J. L. Sambricio, et al. ``Raman spectroscopy of highly pressurized graphene membranes''. {\em Applied Physics Letters}, 108, 221907, (2016).}.

%
%
%
%
%
%

%
%
%

\section{Fit procedures}

Maps reported in Fig.~3a and 3c were obtained by fitting the $G$ region of the Raman spectra (over the range going from $1400\,{\rm cm^{-1}}$ to $1700\,{\rm cm^{-1}}$) using a single lorentzian peak

\begin{align}
A\frac{\left(\Gamma_G/2\right)^2}{\left(\Delta\omega-\Delta\omega_G\right)^2+\left(\Gamma_G/2\right)^2}+B
\end{align}

For samples presenting a clear peak related to disorder, a further lorentzian was added to the fit procedure and subtracted from the dataset. In the $G_\pm$ analysis, two lorentzian peaks were used. In this case, the value of the peak widths $\Gamma_{G_\pm}$ were locked to the value obtained for $\Delta P=0$. Fitting the data with free $\Gamma_{G_\pm}$ did not impact in any significant way the best fit values $\Delta\omega_{G_\pm}$ but yielded an artificially low value of the error on the estimate of the peak positions.

\section{Sample fabrication}

Free-standing graphene regions were obtained using SiN membranes patterned with through holes with various geometries. These were achieved by patterning, using UV lithography and conventional photoresist, the back side SiN with a square with an approximate size of $700\times 700\,{\rm \mu m^2}$. The SiN layer was then removed by dry etching ($200\,{\rm W}$ plasma using CF$_4$ and H$_2$ with a flow of $20$ and $10\,{\rm sccm}$, respectively). In a second patterning, the desired hole geometries are defined on the front side using an aligned e-beam lithographic step and CSAR AR-P6200 resist. After a second dry etching process, the sample is dipped for about $3.5$ hours in a KOH:H$_2$O (1:2) solution heated at $80\,{\rm ^\circ C}$. The top holes in the nitride are created when the SiN membrane is still attached to the bulk of the Si chip and the wet etching of the Si allows a progressive release of the membrane and a relaxation of built-in strain in the SiN. This procedure is crucial to define holes with a geometry which is more complex than a standard circular hole. When holes have a geometry with large strain accumulation points (such as, for instance, in the case of ellipses with a large eccentricity) a direct etching of the suspended SiN can in fact easily lead to a breakdown of the membrane during the nitride etching.

Large monocrystals (typical size $100-200\,{\mu m}$) of  CVD graphene were grown on oxygen-passivated copper foil using an Aixtron BM Pro cold-wall reactor at a pressure of $25\,{\rm mbar}$ and a temperature of $1060\,{\rm^\circ C}$. To allow precise positioning of monocrystals over the elliptical holes, graphene was deposited on SiN substrates using semi-dry transfer, outlined in the following. Graphene was coated with a PMMA support layer  and detached from the Cu growth substrate using electrochemical delamination (``Bubbling transfer''). Utilizing a semi-rigid polyimide frame, a freestanding graphene/PMMA membrane was rinsed in deionized water and dried. Finally, a micromanipulator stage was used to align the graphene/PMMA membrane with the elliptical holes and to attach it to the target substrate.

\end{document}